\documentclass[twocolumn,australian,english,aps,prl,two column,floatfix]{revtex4-1}
\usepackage[T1]{fontenc}
\usepackage[latin9]{inputenc}
\usepackage{bm}
\usepackage{amsmath}
\usepackage{amssymb}
\usepackage{graphicx}

\makeatletter

\@ifundefined{textcolor}{}
{%
\definecolor{BLACK}{gray}{0}
\definecolor{WHITE}{gray}{1}
\definecolor{RED}{rgb}{1,0,0}
\definecolor{GREEN}{rgb}{0,1,0}
\definecolor{BLUE}{rgb}{0,0,1}
\definecolor{CYAN}{cmyk}{1,0,0,0}
\definecolor{MAGENTA}{cmyk}{0,1,0,0}
\definecolor{YELLOW}{cmyk}{0,0,1,0}
}

\makeatother

\usepackage{babel}
\begin{document}

\title{One-dimensional Bose gas dynamics: breather relaxation}

\author{Bogdan Opanchuk and Peter D. Drummond}

\affiliation{Centre for Quantum and Optical Science, Swinburne University of Technology,
Melbourne 3122, Australia}
\begin{abstract}
One-dimensional Bose gases are a useful testing-ground for quantum
dynamics in many-body theory. They allow experimental tests of many-body
theory predictions in an exponentially complex quantum system. Here
we calculate the dynamics of a higher-order soliton in the mesoscopic
case of $N=10^{3}-10^{4}$ particles, giving predictions for quantum
soliton breather relaxation. These quantum predictions use a truncated
Wigner approximation, which is a $1/N$ expansion, in a regime where
other exactly known predictions are recovered to high accuracy. Such
dynamical calculations are testable in forthcoming BEC experiments.
\end{abstract}
\maketitle
Techniques for observing near lossless quantum dynamics have led to
quantitative tests of quantum field dynamics in photonic systems~\cite{Rosenbluh1991,Drummond1993-solitons,Heersink2005,Corney:2006_ManyBodyQD}.
Improvements in ultra-cold quantum gas experiments mean that these
experiments can now also compare first principles calculations of
many-body quantum dynamics with observations~\cite{Egorov_2011}.
The 1D Bose gas, with its well-understood conservation laws~\cite{thacker1981exact}
and exact solutions~\cite{lieb1963exact,mcguire1964study} is an
excellent testing ground for these ideas. Second-order correlations
in thermal equilibrium with repulsive interactions have been predicted~\cite{GangardtPhysRevLett.90.010401,kheruntsyan2003pair,kheruntsyan2005finite,YUROVSKY200861}
and verified experimentally~\cite{tolra2004observation,KinoshitaPhysRevLett.95.190406}.
In these systems, there is evidence of steady-states that do not have
a Gibbs structure~\cite{langen2015experimental,rigol2007relaxation}.
Attractive matter-wave solitons have also been experimentally observed~\cite{khaykovich2002formation,strecker2002formation,medley2014evaporative,mcdonald2014bright,nguyen2014collisions,nguyen2017formation}.

Here we show that the dynamical stability of higher-order matter-wave
solitons prepared by quenching is experimentally testable. Fragmentation
and damping of breathing oscillations~\cite{weiss2016higher,yurovsky2017dissociation}
are predicted to persist even up to a mean particle number of $N=1000$.
These calculations use the truncated Wigner approximation, which is
a $1/N$ expansion~\cite{Graham1973,Drummond1993,Steel1998}. Known
conserved quantities are replicated with high accuracy. This is a
regime accessible to current BEC experiments~\cite{yurovsky2017dissociation,everitt2017observation,nguyen2017formation}.
We show that direct experimental tests of predictions for soliton
fragmentation and center-of-mass dynamics are possible in an exponentially
complex regime where exact calculation is extremely difficult. 

Fragmentation causes a decay in oscillation that is predicted to happen
gradually, without the abrupt changes after a short evolution time
found by variational methods~\cite{streltsov2008formation}. Such
methods are known to disagree with exact COM spreading results~\cite{cosme2016center},
which means that they violate Galilean invariance~\cite{tao2006nonlinear}.
We show that this is because the number of dissociation channels is
much larger than the number of variational modes used in such calculations.
The oscillation decay found here is slower than predicted at very
small particle number~\cite{weiss2016higher}, and also less pronounced
than the predicted fragmentation at small $N$ obtained from exact
analysis~\cite{yurovsky2017dissociation}. However, this difference
is qualitatively consistent with the scaling we find with $N$: where
fragmentation and breather relaxation are reduced as $N$ increases.

Here we investigate the dynamics of a higher-order soliton or breather.
In this case, even more dramatic effects can occur due to quantum
fragmentation. Due to the enormous state-space, direct calculation
with exact eigenstates is not practical in the regime of experimental
interest, with $1000$ particles or more. 

This has been the topic of several publications. The first~\cite{streltsov2008formation}
used a multi-configurational time-dependent Hartree method for bosons
(MCTDHB) approach with $N=1000$ and two spatial modes, predicting
a sudden break-up into a pair of equal size fragments. This calculation
was recently shown to be not fully converged, violating known quantum
center-of-mass (COM) expansion physics~\cite{cosme2016center}. Our
results confirm this earlier analysis. The predictions obtained here
are completely different, with no evidence of a sudden breakup after
a fixed evolution time.

Other approaches have used either exact methods~\cite{yurovsky2017dissociation},
or matrix product states~\cite{weiss2016higher}. These were limited
in number to $N<20$. Here we investigate larger particle numbers,
i.e., $N=10^{3}-10^{4}$, and large numbers of independent modes,
of order $10^{5}$. This is an experimentally realistic regime. Our
calculations preserve all local conservation laws and (nearly) exact
COM dynamics, giving results that are both quantitatively and qualitatively
different to earlier variational studies.

In one dimensional optical or atomic waveguides, a similar Hamiltonian
applies to either massive atomic Bose-Einstein condensate (BEC) experiments
or to photonic experiments, where dispersion gives rise to an effective
mass. If the bosons are confined to a single transverse mode, one
obtains an 1D Bose gas theory, valid for low energies: 
\begin{align}
\hat{H}_{\mathrm{1D}}= & \frac{\hbar^{2}}{2m}\int\hat{\Psi}_{\mathrm{1D}}^{\dagger}H_{1}\hat{\Psi}_{\mathrm{1D}}dr_{3}\nonumber \\
 & +\frac{g_{\mathrm{1D}}}{2}\int\left(\hat{\Psi}_{\mathrm{1D}}^{\dagger}\right)^{2}\hat{\Psi}_{\mathrm{1D}}^{2}dr_{3}\,.
\end{align}
Here, $\bm{r}$ is the spatial coordinate, with a one-dimensional
confinement so the dynamics occur in the $r_{3}$ direction. The mass
is $m$, and for an atomic Bose gas in a parabolic trap one has: 
\begin{align}
H_{1} & =-\hbar^{2}\partial_{3}^{2}/2m+m\omega_{3}^{2}r_{3}^{2}/2\nonumber \\
g_{\mathrm{1D}} & =2\hbar\omega_{\perp}a\,,
\end{align}
where $a$ is the three-dimensional S-wave scattering length, and
the effective transverse trapping frequency of: $\omega_{\perp}=\sqrt{\omega_{1}\omega_{2}}$.
If the system is photonic or polaritonic, as in a fibre optical experiment~\cite{drummond1987quantum,Drummond1993-solitons,Drummond1993},
the relevant parameters come from the dispersion and optical nonlinearity
properties of the fiber. 

This can be transformed to dimensionless form by choosing a length
scale $r_{0}$ and time scale $t_{0}$ such that $r_{0}^{2}=\hbar t_{0}/2m$.
Distance is scaled to so that $z=r_{3}/r_{0}$, and time is scaled
to give a dimensionless time $\tau=t/t_{0}$. The resulting Hamiltonian,
in the form introduced by Lieb and Liniger~\cite{lieb1963exact},
with a dimensionless wave-function $\hat{\psi}=\sqrt{r_{0}}\hat{\Psi}_{\mathrm{1D}}$,
is:
\begin{equation}
\hat{H}=\int dz\left[\hat{\psi}_{,z}^{\dagger}(z)\hat{\psi}_{,z}(z)+C\left(\hat{\psi}^{\dagger}(z)\right)^{2}\hat{\psi}^{2}(z)\right].\label{eq:Hamiltonian}
\end{equation}
We use a subscript to indicate a derivative, so that: 
\begin{equation}
\hat{\psi}_{,z}(z)\equiv\partial_{z}\hat{\psi}(z)\equiv\frac{\partial}{\partial z}\hat{\psi}(z)\,.
\end{equation}

The following relationships exist between the physical and dimensionless
units in the case of a trapped Bose-Einstein condensate~\cite{olshanii1998atomic,kheruntsyan2005finite}:
$\hat{H}=\hat{H}_{1}/E_{0}$, $E_{0}=\hbar/t_{0}=\hbar^{2}/2mr_{0}^{2}$
and $C=mg_{\mathrm{1D}}r_{0}/\hbar^{2}=2m\omega_{\perp}r_{0}a/\hbar$.
 A convenient procedure for solitons is to simply define $r_{0}$
as the characteristic initial dimension, so that $C$ is of the order
of the inverse particle number $N$.

The corresponding dynamical equation is known as the one-dimensional
quantum nonlinear Schrodinger equation. It also describes quantum
photonic propagation in one-dimensional optical fibers~\cite{Carter1987},
under similar conditions of tight transverse confinement. Thus, an
almost identical picture holds for 1D photonic systems~\cite{Carter1987,drummond1987quantum},
except for additional Raman-Brillouin coupling to phonons, owing to
the use of dielectric waveguides~\cite{gordon1986theory,carter1991squeezed}.
This earlier work used phase-space techniques that originate in the
work of Wigner~\cite{Wigner1932} and Glauber~\cite{Glauber1963-states}.
Such predictions have been experimentally verified~\cite{Rosenbluh1991,drummond1987quantum,Corney_2008}.
In both the photonic and atomic experiments, there are additional
dissipative couplings due to linear and nonlinear losses and phase
noise, leading to additional corrections. For simplicity, dissipation
is ignored here, which limits the applicable interaction time. 

The initial quantum states of experimental photonic pulses or BECs
typically has a shot-to-shot randomness in the state preparation that
results in experimental number fluctuations. It is common to have
at least a Poissonian number variance~\cite{chuu2005direct} when
the atom numbers are larger than $10^{3}$. Accordingly, we assume
Poissonian number fluctuations in the calculations given here, in
order to represent typical initial quantum density matrices. The Wigner
distribution $W\left[\psi\right]$ over Wigner fields $\psi$ exists
for any quantum state~\cite{Wigner1932,hillery1984distribution}.
It is not always positive definite. The usual operator time-evolution
equation
\begin{equation}
\frac{d\hat{\psi}}{dt}=-i\left[\hat{H},\hat{\psi}\right],
\end{equation}
where the Hamiltonian $\hat{H}$ is defined by Eq.~(\ref{eq:Hamiltonian}),
can be transformed~\cite{Moyal1949} into a differential equation
for $W\left[\psi\right]$, typically with third or higher order derivatives.
After truncation of third order derivatives~\cite{Graham1973}, which
are the highest order terms in a $1/N$ expansion for $N$ particles,
one obtains a second order Fokker-Plank equation for $W\left[\psi\right]$.
This is an approximate functional differential equation for a probability
distribution over Wigner fields. 

When the evolution is unitary, this results in a partial differential
equation for phase-space variables using well-known procedures~\cite{Steel1998,opanchuk2013functional,Drummond1993}.
The resulting equation for the Wigner field $\psi$, is:
\begin{equation}
\frac{d\psi}{dt}=i\nabla^{2}\psi-2iC\psi\left(|\psi^{2}|-1/\Delta z\right),
\end{equation}
where $\Delta z$ is the lattice spacing or inverse momentum cutoff.
Quantum noise is present in the initial conditions. We start from
a state with Poissonian number distribution, which is equivalent to
a coherent state:
\begin{equation}
\hat{\rho}\left(t=0\right)=|\alpha\left(z\right)\rangle\langle\alpha\left(z\right)|,
\end{equation}
where $|\alpha\left(z\right)|^{2}=n\left(z\right)$. In the Wigner
representation this is exactly represented by an ensemble of fields
$\psi\left(z\right)$ with initial quantum noise $\eta_{k}$, with
\begin{equation}
\psi\left(z\right)=\sqrt{n\left(z\right)}+\frac{1}{\sqrt{2}}\sum_{k}\frac{1}{\sqrt{L}}\eta_{k}e^{ikz}.
\end{equation}
Here $\eta_{k}$ are complex random numbers correlated as $\langle\eta_{k}\eta_{k^{\prime}}^{*}\rangle=\delta_{kk^{\prime}}$,
$\langle\eta_{k}\eta_{k^{\prime}}\rangle=0$. The functional integration
over the Wigner distribution is performed by generating multiple random
initial states and using them to seed independent integrations of
the PDE\@. This results in a large number, $N_{\mathrm{s}}$, of
independent field modes~\textemdash{} each evolving in time with
equal probability.

The Wigner phase-space method generates a direct representation of
symmetrically ordered quantum observables. To obtain the usual normally-ordered
quantum observables, one must transform the results of a Wigner calculation
from a symmetrically ordered to a normally ordered form. This also
removes the divergence of symmetrically-ordered observables at large
momentum cutoff. The expectation values of symmetrically ordered operator
expressions can be obtained by integrating this equation over multiple
independent trajectories to produce a set of values $\psi^{(j)}$
and averaging over a corresponding function of these values. 

There is an approximate equality between symmetrically ordered quantum
averages and Wigner averages, where the $N$-dependent truncation
error depends on the evaluated operator~\cite{kinsler1993limits,Kinsler:1996}:
\begin{equation}
\left\langle \left\{ \hat{O}\left(\hat{\psi},\hat{\psi}^{\dagger}\right)\right\} \right\rangle \approx\left\langle \left\{ \hat{O}\right\} \right\rangle _{W}=\langle O\rangle_{W}.
\end{equation}

We consider a quantum dynamical experiment where an initial state
is prepared and then evolved in time. The initial state is a Poissonian
mixture of uncorrelated particles with mean value $N=10^{3}-10^{4}$
in a localized spatial mode. The equivalent coherent state has the
classical soliton shape that occurs with some small initial coupling
of $C_{i}=-2/N$, with $r_{0}$ as the characteristic initial size,
so that in dimensionless units, $\alpha(z)=\sqrt{N/2}\,\mathrm{sech}(z)$.

\begin{figure}
\includegraphics{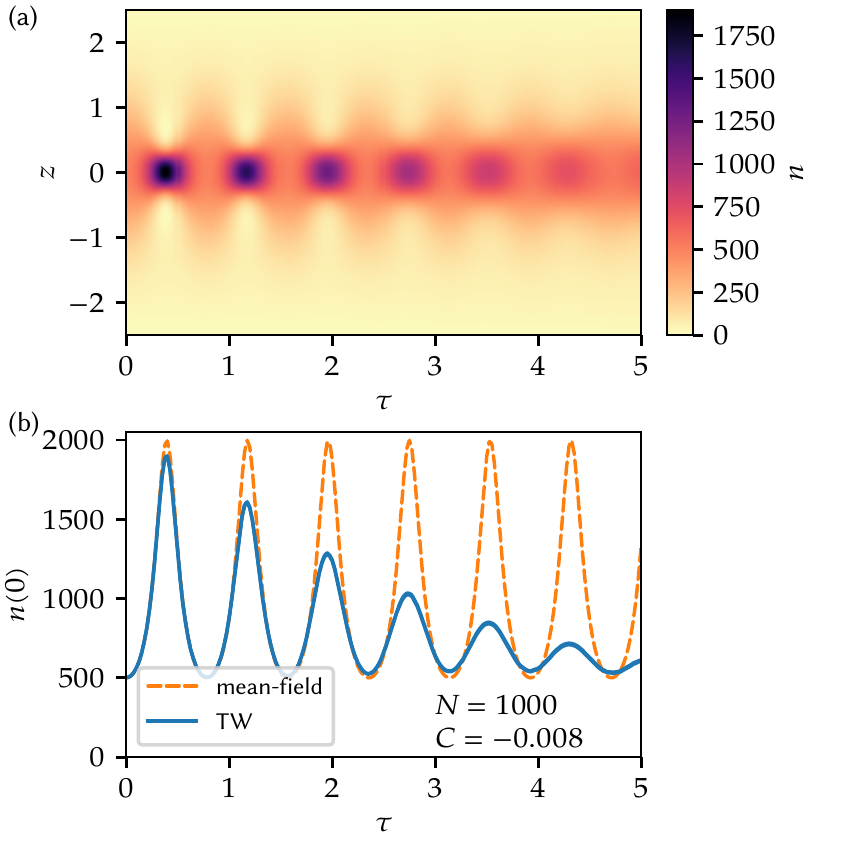}

\caption{\label{fig:Density-N1k}Density near the centre of the simulation
area (a) and at $z=0$ (b) over time. Simulation with $N=10^{3}$,
$C=-8\times10^{-3}$, $M=512$, $L=20$, $10^{5}$ trajectories, $10^{5}$
time steps. The area between the simulation curves (solid blue lines)
denotes the estimated sampling error. The result of the mean-field
simulation (dashed orange lines) are shown for comparison. The time-step
errors are smaller than the line thickness and are not shown on the
graph.}
\end{figure}

\begin{figure}
\includegraphics{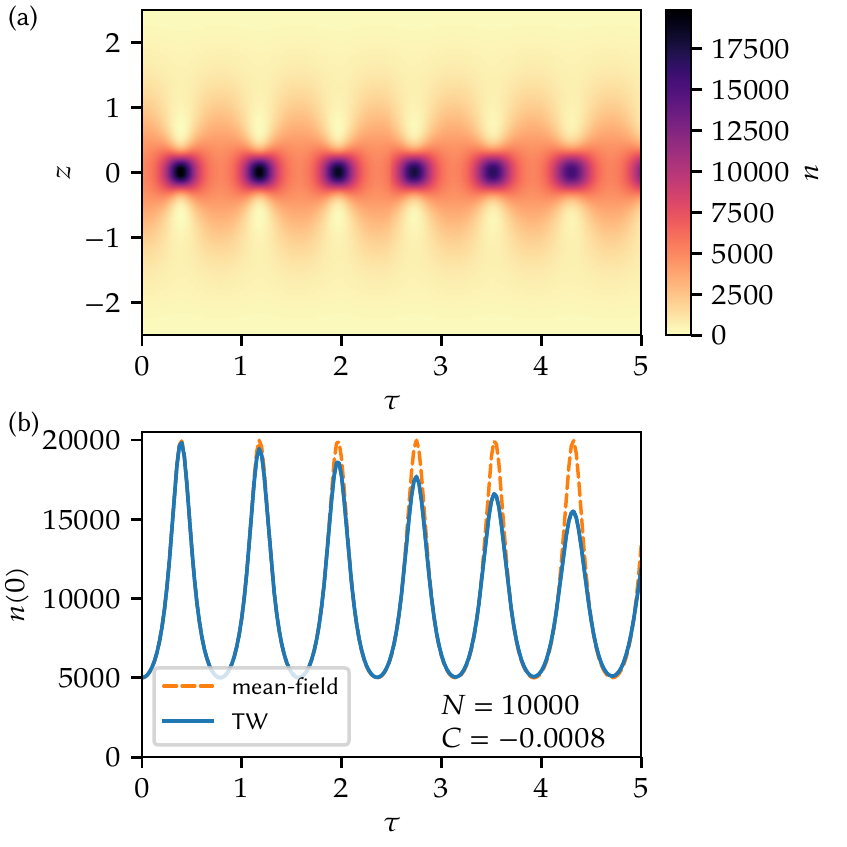}

\caption{\label{fig:Density-N10k}Density near the centre of the simulation
area (a) and at $z=0$ (b) over time. Simulation with $N=10^{4}$,
$C=-8\times10^{-4}$, other properties as in Fig.~\ref{fig:Density-N1k}.}
\end{figure}

This corresponds to an ultra-cold atomic Bose gas experiment, with
a BEC initially trapped in a localized state with no interactions.
At time $t=0$, the interaction Hamiltonian is turned on to a larger
value of $C_{f}=-8/N$, allowing particles to interact and forming
a breather, a higher-order oscillating soliton. The resulting density
profile, $\left\langle \hat{n}\left(z\right)\right\rangle =\left\langle \hat{\psi}^{\dagger}(z)\hat{\psi}(z)\right\rangle $,
is shown in Fig.~\ref{fig:Density-N1k} for $N=1000$ and in Fig.~\ref{fig:Density-N10k}
for $N=10000$. The result of the initial condition is that a high-order
soliton or breather is formed~\cite{wai1986nonlinear}, with a characteristic
period of $\tau_{b}=\pi/4$. Our numerical results show characteristic
breathing oscillations such that the mean breather amplitude decays
with time. 

This simulation is similar to related experimental proposals of first
creating a fundamental soliton at weak coupling, then suddenly increasing
the coupling strength. The coupling change would be caused by either
a pulse entering a fiber in a photonic experiment, or else a change
in a tunable Feshbach resonance in an atomic system. A number of different
theoretical methods~\cite{streltsov2008formation,weiss2016higher,yurovsky2017dissociation}
have been used to analyze this type of proposed experiment, making
it of topical interest. The present protocol employs a localized non-interacting
BEC as the initial state, following earlier proposals~\cite{streltsov2008formation,cosme2016center}.
The timescales and numbers used are within the general parameter range
achievable with current $^{7}$Li~\cite{nguyen2017formation} and
$^{85}$Rb~\cite{everitt2017observation} ultra-cold atomic physics
experiments.

The simulation is sensitive to the selected spatial and momentum grids.
The spatial grid must be symmetrical around 0 and have a point at
$z=0$, or else the decay happens on a faster scale, since there is
insufficient lattice resolution for spatial convergence. The momentum
grid should ideally be symmetrical around 0, which can be achieved
by using a pair of position- and momentum-dependent coefficients applied
before and after the Fourier transform. If this condition is not satisfied,
the unbalanced high-momentum components of the noise lead to numerical
errors. A finite lattice was used with periodic boundary conditions
at $z=\pm L/2$. Results were obtained using a public domain stochastic
partial differential equation code~\cite{Kiesewetter2016xspde} with
a fourth-order Runge-Kutta interaction picture algorithm~\cite{Caradoc-Davies2000},
then cross-checked with a larger number of samples using an open source
graphical processor unit (GPU) code.

The initial density matrix used here is a random phase mixture of
coherent states. This is exactly equivalent to a Poissonian mixture
of initial pure number states in a single spatial mode, chosen as
$u=\mathrm{sech}\left(z\right)/\sqrt{2}$, similar to previous investigations~\cite{streltsov2008formation,cosme2016center}.
Since the measurements phase-independent, only a single phase in the
mixture is calculated. Averaging over phases would produce identical
results in every input phase.In the present examples, the initial
boson number is $N=N\pm\sqrt{N}$, where $N=10^{3}-10^{4}$. The number
standard deviation is $\pm\sqrt{N}$, or $\pm1\%-3.2\%$, which is
typical for these types of experiment.

Convergence tests were carried out with the four exact conservation
laws, $\hat{N}$, $\hat{P}$, $\hat{H}$, $\hat{H}_{3}$~\cite{davies1990higher},
and with exact COM expansion predictions~\cite{kohn1961cyclotron,vaughan2007quantum}.
All agreed with the predicted conserved behavior, apart from small
errors of size $N^{-3/2}$. The comparison with these tests will be
reported in detail elsewhere. Truncated Wigner methods can have a
growing truncation error with time~\cite{Sinatra2002,Deuar2007};
however, earlier variational results were not able to satisfy these
tests~\cite{cosme2016center}. The main issue is whether the breather
behaves classically, or whether the oscillations are damped owing
to quantum fragmentation of the higher-order soliton. This problem
is extremely challenging in quantum many-body theory, as it involves
exponentially many eigenstates. As can be seen by the results given
here, in the TW approximation the oscillations are predicted to decay
gradually, without sudden fragmentation as predicted using variational
methods~\cite{streltsov2008formation}. 

\begin{figure}
\includegraphics{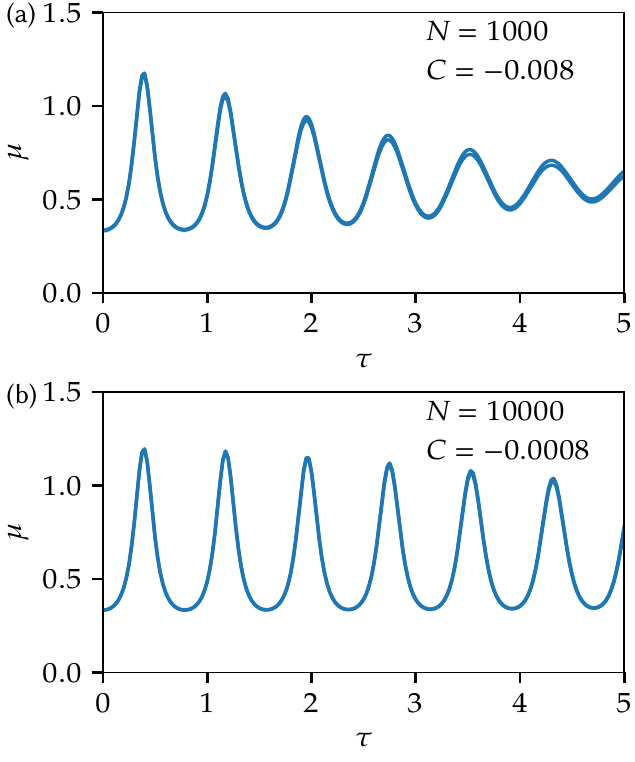}

\caption{\label{fig:Correlation--over}Correlation $\mu$ over time. Simulation
with $N=10^{3}$, $C=-8\times10^{-3}$ (a) and $N=10^{4}$, $C=-8\times10^{-4}$
(b), $M=512$, $L=20$, $10^{5}$ trajectories, $10^{5}$ time steps.
The area between the simulation curves (solid blue lines) denotes
the estimated sampling error. The time-step errors are smaller than
the line thickness and are not shown on the graph.}
\end{figure}

Since the center-of mass position is known to spread, one may expect
that the on-axis density plotted in Fig.~\ref{fig:Density-N1k} and
Fig.~\ref{fig:Density-N10k} might decay purely due to the quantum
uncertainty in the final position. Therefore, in Fig.~\ref{fig:Correlation--over},
we introduce the dimensionless Glauber second order correlation function,
$G^{(2)}\left(z_{1},z_{2}\right)=\left\langle \hat{\psi}^{\dagger}\left(z_{1}\right)\hat{\psi}^{\dagger}\left(z_{2}\right)\hat{\psi}\left(z_{2}\right)\hat{\psi}\left(z_{1}\right)\right\rangle $,
and investigate the integrated correlation:
\begin{equation}
\mu=\int G^{(2)}\left(z,z\right)dz/N^{2}.
\end{equation}

\begin{figure}
\includegraphics{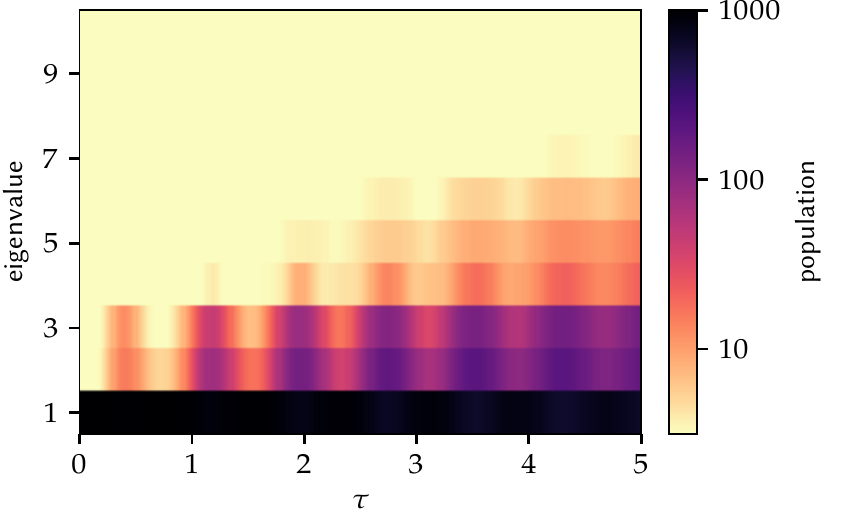}

\caption{\label{fig:Eigenvalues}Eigenvalues of the correlation function $G^{(1)}$
over time. Simulation with $N=10^{3}$, $C=-8\times10^{-3}$ $M=512$,
$L=20$, $10^{3}$ trajectories, $10^{5}$ time steps. The graph shows
increasing fragmentation with time.}
\end{figure}

This integrated correlation function measures the ``peakedness''
of a spatial distribution, in a way that is independent of the location
of the peak. This also decays, although not as strongly as the on-axis
density. We conclude that the breather appears to gradually radiate
or fragment due to quantum effects with increasing similarity to mean
field \foreignlanguage{australian}{behaviour} as $N\rightarrow\infty$.
This is confirmed by an eigenvalue analysis of the first order correlation
function, $G^{(1)}\left(z_{1},z_{2}\right)=\left\langle \hat{\psi}^{\dagger}\left(z_{1}\right)\hat{\psi}\left(z_{2}\right)\right\rangle $.
The definition of a Bose condensate is that it has a macroscopic occupation~\cite{penrose1956bose}of
a single eigenmode of $G^{(1)}$. The transition to a partially fragmented
BEC is illustrated in Fig.~\ref{fig:Eigenvalues}, which shows that
six modes dynamically evolve to $>1\%$ occupation by $\tau=5$. This
cannot be treated accurately by variational calculations with fewer
modes~\cite{cosme2016center}.

In summary, our results predict continuous quantum fragmentation of
higher-order soliton breathers at particle numbers of $N=1000$, with
results closer to mean field predictions at $N=10000$. This is readily
testable in BEC experiments.
\begin{acknowledgments}
We would like to acknowledge helpful discussions with J. Brand, J.
Cosme, R. Hulet, B. Malomed, M. Olshanii and L. Carr. This work was
performed in part at Aspen Center for Physics, which is supported
by National Science Foundation grant PHY-1607611.
\end{acknowledgments}

\bibliographystyle{apsrev4-1}
\bibliography{1DSchro}

\end{document}